\documentclass[12pt,a4paper]{article}
\begin{document}

\title{D-Instantons in Non-Critical Open String Theory}
\author{\underline{Rui Neves}\thanks{Research supported by
J.N.I.C.T's PRAXIS XXI
PhD fellowship BD/2828/93-RM.}\\
\footnotesize{Department of Mathematical Sciences,
University of Durham, Science Laboratories,}\\
\footnotesize{South Road, Durham DH1 3LE, United Kingdom}\\
\footnotesize{R.G.M.Neves{\it @}Durham.ac.uk}}
\maketitle

\begin{abstract}
We show that the strength of the leading non-perturbative effects in
non-critical string theory is of the order $e^{-O(1/{\beta_{st}})}$.
We show how this restricts the space of consistent theories. We also
identify non-critical one dimensional D-instantons as dynamical
objects which exchange closed string states and calculate the order
of their size. PACS codes: 11.25-w, 11.25.Pm, 11.25.Hf, 11.25.Sq.
\end{abstract}

\section{Introduction}

Over the years string theory has been establishing itself as the leading
candidate to a unified description of particle physics and gravity. The
latest and exciting development has been the rediscovery of string
duality \cite{JP,CJP}. Particularly fascinating is the connection
between this
symmetry and new extended objects like the D-branes, which has given
strength to the idea that boundaries and dual membranes play a significant
role in the
understanding of the non-perturbative phase of the dynamics of strings
\cite{DLP,JPO}. In this letter we are interested in dual membranes in
non-critical open string theory. We consider Polyakov's
model and work within the DDK approach to the conformally invariant
Liouville theory \cite{DDK, MN}. We discuss the non-critical
D-instanton in
the limiting case of a one dimensional target space and show that the
strength of its associated leading stringy
non-perturbative effects is of the order $e^{-O(1/{\beta_{st}})}$, where
$\beta_{st}$ is the string coupling constant. This naturally mimics
the result obtained in the $26$ dimensional critical theory
and in the matrix
models \cite{JPO, SS} since the weight of holes in the world-sheet only
depends on the topology. However we find that not all of the non-critical
theories allowed by perturbative Weyl invariance are consistent. These
theories are characterised by different positive values of the
renormalised Liouville cosmological constants $\lambda_2$ and $\mu_2$,
respectively associated with the renormalisation counterterms in
the length of the boundary and in the area of the world-sheet. We show
that only
those which satisfy ${\lambda_2}\geq{\mu_2}$ with ${\lambda_2}>0$,
${\mu_2}\geq 0$ or those where ${\lambda_2}=0$, ${\mu_2}\geq 0$ lead to
acceptable non-perturbative effects.
We also consider the D-instanton exchange of closed string states
and show that the size of the D-instanton is
of the order of $\sqrt{\alpha'}/\ln(1/{\lambda_2})$ for small
${\lambda_2}>0$, ${\mu_2}\geq 0$ or of the order of
$\sqrt{\alpha'}/\ln(1/{\mu_2^2})$ for ${\lambda_2}=0$ and small
${\mu_2}>0$. Above $\alpha'$ is related to the string tension
$\mbox{T}=1/(2\pi\alpha')$.
The extension of the analysis to the $c\leq 1$ minimal boundary
conformal models is briefly discussed.

\section{Non-critical D-instantons}

For a string in a non-critical
target space we have to take into the picture the non-linear
dynamics of the conformally invariant path integral Liouville theory.
Here the boundary conditions we may impose on the
Liouville mode are constrained by quantum Weyl invariance. So while the
Dirichlet boundary conditions lead to a
discontinuity in the metric as the boundary is approached, the Neumann and
free boundary conditions both allow a fully smooth and Weyl invariant
theory \cite{MN}. For simplicity we use Neumann
boundary conditions on the Liouville mode.

We start
with the disc non-critical partition function with the case of the
D-instanton in mind \cite{JPO}. Consider the dual of an open bosonic
string theory
with $k$ dimensions compactified on a torus. The dual string field
satisfies homogeneous Dirichlet boundary conditions on the compact
dimensions ${Y^i}{|_B}=0$, $i=d-k+1,\ldots,d$ and Neumann boundary
conditions on the non-compact dimensions ${\partial_{\tilde{n}}}{Y^m}=0$,
$m=1,\ldots,d-k$. Omiting the counterterms the partition function is
given by

\begin{equation}
Z={\beta_{st}^{-{\chi_o}}}\int{\mathcal{D}_{\tilde{g}}}(Y,\tilde{g})
\exp\left(-{{\mbox{T}}
\over{2}}
\int{d^2}\xi\sqrt{\tilde{g}}Y\cdot\tilde{\Delta}Y\right),
\end{equation}
where for the disc ${\chi_o}=1$. Integrating the matter fields and the
reparametrisation ghosts we find after the DDK renormalisation

\begin{equation}
Z={\beta_{st}^{-{\chi_o}}}{{\left({{\mbox{T}}\over{2\pi}}\right)}^
{(d-k)/2}}
{{\left({{{\mbox{Det}_N}'
\hat{\Delta}}\over
{\int{d^2}\xi\sqrt{\hat{g}}}}\right)}^{-(d-k)/2}}
{{({\mbox{Det}_D}\hat{\Delta})}^{-k/2}}{{\sqrt{\mbox{Det}'
{\hat{P}^\dagger}\hat{P}}}\over{\mbox{Vol}(\mbox{CKV})}}
\int{\mathcal{D}_{\hat{g}}}\phi{e^{-{S_L}[\phi,\hat{g}]}},
\end{equation}
where the renormalised Liouville action is

\begin{equation}
{S_L}\left[\phi,\hat{g}\right]={1\over{8\pi}}\int{d^2}\xi
\sqrt{\hat{g}}\left({1\over{2}}\phi\hat{\Delta}\phi+Q\hat{R}\phi\right)
+{Q\over{4\pi}}
\oint{d}\hat{s}{k_{\hat{g}}}\phi
+2{\mu_2^2}\int{d^2}\xi
\sqrt{\hat{g}}{e^{\alpha\phi}}
+{\lambda_2}\oint{d}\hat{s}{e^{\alpha
\phi/2}}.
\end{equation}
For Weyl invariance we have
$Q=\pm\sqrt{(25-d)/6}$ and for the Liouville field renormalisation
${\alpha_{\pm}}=(1/2)(Q\pm\sqrt{{Q^2}-4})$ \cite{DDK, MN}.

To integrate the Liouville mode we separate out its zero mode
\cite{ADH, DFK}.
To do so we take the semi-classical branch of $\alpha$, $\alpha_-$,
which in
what follows will be denoted as $\alpha$. We then write
$\phi={\phi_0}+\bar{\phi}$, where $\int{d^2}\xi\sqrt{\hat{g}}
\bar{\phi}=0$. To deal with the boundary terms we have to use
the Mellin transform \cite{Bat}. Then the partition function is

\[
Z={\beta_{st}^{-{\chi_o}}}{{\left({{\mbox{T}}\over{2\pi}}\right)}^
{(d-k)/2}}
{{\left({{{\mbox{Det}_N}'
\hat{\Delta}}\over
{\int{d^2}\xi\sqrt{\hat{g}}}}\right)}^{-(d-k)/2}}
{{({\mbox{Det}_D}\hat{\Delta})}^{-k/2}}{{\sqrt{\mbox{Det}'
{\hat{P}^\dagger}\hat{P}}}\over{\mbox{Vol}(\mbox{CKV})}}
\int{\mathcal{D}_{\hat{g}}}\bar{\phi}
{{2^{s-1}}\over{\pi\alpha}}
\]
\[
\times
{{\left(\int{d^2}\xi
\sqrt{\hat{g}}\right)}^{1/2}}
{{\left(2{\mu_2^2}\int{d^2}\xi
\sqrt{\hat{g}}{e^{\alpha\bar{\phi}}}\right)}^s}
\Gamma(-2s)\exp\left[
{{{\lambda_2^2}{{\left(\oint d\hat{s}{e^{\alpha\bar{\phi}/2}}
\right)}^2}}\over{16{\mu_2^2}\int{d^2}\xi\sqrt{\hat{g}}
{e^{\alpha\bar{\phi}}}}}\right]
\]
\begin{equation}
\times
{D_{2s}}\left[
{{{\lambda_2}\oint d\hat{s}{e^{\alpha\bar{\phi}/2}}}\over{
2{\mu_2}{{\left(\int{d^2}\xi\sqrt{\hat{g}}
{e^{\alpha\bar{\phi}}}\right)}^{1/2}}}}
\right]{e^{-{S_L^0}[\bar{\phi},\hat{g}]}},
\end{equation}
where $\alpha s=Q{\chi_o}/2$, ${D_{2s}}(z)$ are the parabolic cylinder
functions \cite{Bat} and

\begin{equation}
{S_L^0}[\bar{\phi},\hat{g}]={1\over{8\pi}}\int{d^2}\xi
\sqrt{\hat{g}}\left({1\over{2}}\bar{\phi}\hat{\Delta}\bar{\phi}+
Q\hat{R}\bar{\phi}\right)
+{Q\over{4\pi}}
\oint{d}\hat{s}{k_{\hat{g}}}\bar{\phi}.
\end{equation}

Take the case where $k=d$. Following Polchinski \cite{JPO}
the partition function on the disc is
to be interpreted as minus the D-instanton action, which appears in the
one D-instanton amplitude ${A_1}\sim{e^{Z}}$. In the critical dimension
this only gives the right weight $e^{-O(1/{\beta_{st}})}$
because of the negative value of the
renormalised M\"obius volume \cite{LP}. Here we have the additional
problem associated with the multiple scattering of the Liouville
vertex operators. Let us try to calculate the partition function in the
limiting case $d=1$. In this case $Q=2$, $\alpha=1$ lead to $s=1$.
Then using ${2^s}{e^{{z^2}/4}}{D_{2s}}(z)={H_{2s}}(z)$ \cite{Bat}, where
${H_{2s}}(z)$ is the Hermite polynomial of degree $2s$, we get
${H_2}(z)=4{z^2}-2$. Integrating over $\bar{\phi}$ we find on the upper
half-plane

\[
Z={{\Gamma(-2)}\over{2\pi{\beta_{st}}}}{{
\left({\mbox{Det}_D}\hat
{\Delta}\right)}^{-1/2}}{{\left({{{\mbox{Det}_N}\hat
{\Delta}}\over{\int{d^2}\xi\sqrt{\hat{g}}}}\right)}^{-1/2}}
{{\sqrt{\mbox{Det}'
{\hat{P}^\dagger}\hat{P}}}\over{\mbox{Vol}(\mbox{CKV})}}
{\varepsilon^{-6}}{\Lambda^8}
\]
\begin{equation}
\times\left[{\lambda_2^2}{\int_{-\infty}^{+\infty}}dxdy{{(x-y)}^{-2}}-
{{\mu_2^2}\over{2}}{\int_{-\infty}^{+\infty}}dxdy{y^{-2}}\right].
\end{equation}
Here $\varepsilon\to 0$ is a short distance cutoff which comes from the
Green's functions calculated at equal points. We have also introduced
the background gravity charge in the boundary at $\Lambda\to\infty$.
To factor out the divergence
associated with the M\"obius invariance, we fix the bulk vertex
operator on the center of the unit disc and
the two boundary operators on $0$ and $\pi/2$. Then using the M\"obius
volume for an oriented theory \cite{LP} we find

\begin{equation}
Z=-{{\left({{3}\over{16\pi}}\right)}^{3/2}}
{4\over{\pi\sqrt{\pi}{\beta_{st}}}}\left({\lambda_2^2}-{\mu_2^2}\right)
\sqrt{Q_S^2},\label{1}
\end{equation}
where ${Q_S^2}={{\left({\mbox{Det}_S}'\hat
{\Delta}\right)}^{-1}}\sqrt{{\mbox{Det}_S}'
{\hat{P}^\dagger}\hat{P}}$ refer to the determinants on the Riemann
sphere and now $\lambda_2$ and
$\mu_2$ are taken as dimensionless
constants.

Now, for any $s$ we may
always absorb $\lambda_2^{2s}$ into $\beta_{st}$ and so obtain
amplitudes which only depend on $\beta_{st}$ and on the ratio
${\mu_2}/{\lambda_2}$. Within the DDK approach the
one dimensional non-critical string theory can be viewed as a two
dimensional critical theory defined in a consistent background given
by the matter and Liouville systems \cite{DFK}. So the ratio of the
Liouville couplings defines different possible backgrounds for the
critical string theory and so different non-critical theories.
{}From perturbative quantum Weyl invariance alone all of them are
allowed.
However, since the zeta function regularisation of the determinants
always leads to positive values, Eq. (\ref{1}) shows that
only for ${\lambda_2}>{\mu_2}$ we can find
non-critical stringy non-perturbative effects
of the order $e^{-O(1/{\beta_{st}})}$ which are in agreement
with the result found in the critical string theory and in the matrix
models \cite{JPO, SS}. If ${\lambda_2}={\mu_2}$ we find no
non-perturbative
effects. If ${\lambda_2}<{\mu_2}$ the theory is inconsistent.

Note that Eq. (\ref{1}) is only valid if ${\lambda_2}>0,
{\mu_2}\geq 0$. If we take ${\lambda_2}=0, {\mu_2}\geq 0$ we just have to
go back to the integral of the Liouville zero mode $\phi_0$ to see that
in this case it does not bring an extra minus sign to $Z$. So, for
${\lambda_2}=0, {\mu_2}\geq 0$ we also find the right weight
$e^{-O(1/{\beta_{st}})}$ due to the negative value of the renormalised
M\"obius volume.

We may also analyse the range $d<1$. This
corresponds to the coupling to 2D quantum gravity of the $c<1$ minimal
boundary conformal field theories \cite{MN, JC, JS}. In this case $d=c=1-
6{{(\beta-1/\beta)}^2}$. Since the background gravity charge only
depends on the world-sheet topology we still find $\alpha s=Q{\chi_o}/2$,
where $Q=\beta+1/\beta$ and $\alpha={\alpha_+}=\beta$. Then on the disc
we find $s=(1+{\beta^2})/(2{\beta^2})$ which is a rational number because
${\beta^2}=(2+k')/(2+k)$, where $k,k'$ are positive
integers. To see if the analysis of the limiting case $d=1$ still holds
for general $k,k'$ we would have to get involved
with the intricate dynamics of the multiple Liouville scattering where
$s$ does not have to be an integer \cite{ADH, DFK}. However if
the boundary dominates the bulk, ${\lambda_2}>{\mu_2}$, the
stringy non-perturbative effects should still be of the
order $e^{-O(1/{\beta_{st}})}$. If just one of the Liouville couplings
is zero the same result is to be expected. These remarks are based on a
possible Coulomb gas representation of the minimal boundary conformal
field theories \cite{MN, JS}. They suggest that the strength of the
non-perturbative
effects is of the order $e^{-O(1/{\beta_{st}})}$. A precise definition
of the effects needs a careful analysis of the partition function on
this type of models. This is beyond the scope of this work and we hope
to discuss the matter in a future paper.

\section{The size of non-critical D-instantons}

To look for D-branes we may also consider the case where there is an
exchange of closed string states while they are separated by a
distance ${\mathcal{Y}^i}{\mathcal{Y}_i}$ \cite{CJP}. The diagram for
such an exchange is an open string annulus. Consider $p+1$
uncompactified dimensions $m=0,\ldots,p$ and $d-p-1$
compact ones. According to T-duality we use Neumann boundary
conditions for the $p+1$ D-brane coordinates and Dirichlet boundary
conditions on the other compact coordinates. We consider that
${Y^i}{|_{B_1}}=-{\mathcal{Y}^i}/2$ and
${Y^i}{|_{B_2}}={\mathcal{Y}^i}/2$, where ${\mathcal{Y}^i}\geq 0$, for all
$i=p+1,\ldots,d$.

As usual the partition function is
$Z={Z_{nc}}{Z_c}$,
where $Z_{nc}$, $Z_c$ are respectively the partition function on the
non-compact and compact space. To integrate on the compact coordinates
we have to factorise the classical solution ${Y^i}={Y_c^i}+{\bar{Y}^i}$,
where ${\bar{Y}^i}$ satisfies homogeneous Dirichlet conditions on both
boundaries and $\tilde{\Delta}{Y_c^i}=0$,
${Y_c^i}{|_{B_1}}=-{\mathcal{Y}^i}/2$, ${Y_c^i}{|_{B_2}}={\mathcal{Y}^i}/2$.

Then integrating the matter and the reparametrisation ghosts fields
we find

\[
Z={V_{p+1}}{{\left({{\mbox{T}}\over{2\pi}}\right)}^{(p+1)/2}}
\int d\tau{{\mbox{det}(\psi,B)}\over{\sqrt{\mbox{det}\hat{(B,B)}}}}
{{\sqrt{\mbox{Det}'{\hat{P}^\dagger}\hat{P}}}\over{\mbox{Vol}(\mbox{CKV})
}}{{\left({{{\mbox{Det}_N}'\hat{\Delta}}\over{\int{d^2}\xi\sqrt{\hat{g}}}}
\right)}^{-(p+1)/2}}
\]
\begin{equation}
\times{{\left({\mbox{Det}_D}\hat{\Delta}\right)}^
{-(d-p-1)/2}}
\exp\left(-{{\mbox{T}}\over{2}}\int{d^2}\xi\sqrt{\hat{g}}
{\hat{g}^{ab}}{\partial_a}{Y_c^i}{\partial_b}{Y_{ci}}\right)\int
{\mathcal{D}_{\tilde{g}}}\varphi{e^{-{S_L}[\varphi,\hat{g}]}},
\end{equation}
where $V_{p+1}$ is the D-brane world volume and $\tau$ is the modulus of
the annulus.
In the case of the annulus ${\chi_o}=0$. So we can choose $\hat{R}=
{k_{\hat{g}}}=0$ and there is no background gravity
charge. This means that we do not have to deal with the Liouville
amplitudes
because $s=0$. After the DDK renormalisation we can integrate $\phi$
with Neumann boundary conditions to obtain

\[
Z={{V_{p+1}\Gamma(0)}\over{2\pi\alpha}}{{\left({{\mbox{T}}\over{2\pi}}
\right)}^{(p+1)/2}}
\int d\tau{{\mbox{det}(\psi,B)}\over{\sqrt{\mbox{det}\hat{(B,B)}}}}
{{\sqrt{\mbox{Det}'{\hat{P}^\dagger}\hat{P}}}\over{\mbox{Vol}(\mbox{CKV})
}}{{\left({{{\mbox{Det}_N}'\hat{\Delta}}\over{\int{d^2}\xi\sqrt{\hat{g}}}}
\right)}^{-(p+2)/2}}
\]
\begin{equation}
\times{{\left({\mbox{Det}_D}\hat{\Delta}\right)}^
{-(d-p-1)/2}}
\exp\left(-{{\mbox{T}}\over{2}}\int{d^2}\xi\sqrt{\hat{g}}
{\hat{g}^{ab}}{\partial_a}{Y_c^i}{\partial_b}{Y_{ci}}\right).
\end{equation}

In this formula it is clear that the only effect of the Liouville mode
within the DDK approach is to introduce just another dimension into
the problem and so
the calculation can be immediately carried out
by following the steps already taken in the critical dimension
\cite{CJP, CMNP}. To describe the annulus we will choose the parameter
domain to be a square $\{0\leq{\xi^a}\leq 1\}$ of area $\tau$. Since we
consider $\xi^1$ as the periodic coordinate we have
${Y_c^i}={\mathcal{Y}^i}{\xi^2}-{\mathcal{Y}^i}/2$. We obtain

\begin{equation}
Z={V_{p+1}}\Gamma(0){{(2\pi{\alpha^2}\mbox{T})}^{-1/2}}
\int{{d\tau}\over{\tau}}
{{\left({{\mbox{T}}\over{4\pi\tau}}\right)}^{(p+2)/2}}
\exp\left(-\mbox{T}{\mathcal{Y}^2}\tau\right){{|\eta(i\tau)|}^{1-d}},
\end{equation}
where $\eta$ is the Dedekind $\eta$ function.

{}From the asymptotics $\tau\to+\infty$ we use
$|\eta(i/\tau)|=\sqrt{\tau}
|\eta(i\tau)|$ to get the asymptotics for $\tau\to 0$ which are
dominated by the lightest closed string states. We then get

\[
Z={V_{p+1}}\Gamma(0){{(2\pi{\alpha^2}\mbox{T})}^{-1/2}}
\int{{d\tau}\over{\tau}}
{{\left({{\mbox{T}}\over{4\pi\tau}}\right)}^{(p+2)/2}}
\exp\left(-\mbox{T}{\mathcal{Y}^2}\tau\right){\tau^{(d-1)/2}}\Big[
{e^{(d-1)\pi/(12\tau)}}
\]
\begin{equation}
+(d-1){e^{(d-25)\pi/(12\tau)}}+\cdots\Big].
\end{equation}
So we find a tower of massive poles for $d<1$ and one massless
tachyon pole for $d=1$. This massless tachyon pole corresponds to the
massless tachyon in the two dimensional critical theory where
the Liouville mode is interpreted as Euclidean time \cite{DFK}.
If we consider a
Coulomb gas representation
of the $c\leq 1$ minimal boundary conformal field theories it is clear
that we get exactly the same result as for the string because $s=0$.
However this only
confirms the suggestion that the order of magnitude of the
non-perturbative effects
should be the same as that of the string. For $p=-1$ the massless pole
contribution is

\begin{equation}
{Z_p}={{\Gamma(0)}\over{2\pi\sqrt{2}}}\int d\tau{\tau^{-3/2}}
\exp\left(-
\mbox{T}{\mathcal{Y}^2}\tau\right).
\end{equation}
This can be integrated using the Gamma function giving

\begin{equation}
{Z_p}=-{{\Gamma(0)}\over{\sqrt{2\pi}}}{{\mbox{T}}^{1/2}}\mathcal{Y}.
\end{equation}

Let us now regularise the divergence associated with the above
Gamma function \cite{DFK}. If ${\lambda_2}>0$ we use the KPZ scaling on
the disc for fixed boundary length $L$ \cite{MN} and write

\begin{equation}
{Z_p}=-\ln(1/{\lambda_2}){{\left({{\mbox{T}}\over{2\pi}}\right)}^{1/2}}
\mathcal{Y}.
\end{equation}
which diverges like $\ln(1/{\lambda_2})$ as the now dimensionless
${\lambda_2}\to 0$. If ${\lambda_2}=0$ we use the
KPZ scaling with fixed area $A$ \cite{MN} to get a similar divergence
with a dimensionless $\mu_2^2$. These are the divergences which correspond
to that already known to appear in the matrix models \cite{GK, BK}.
Note that
$\ln(1/{\lambda_2})$, $\ln(1/{\mu_2^2})$ with dimensionful $\lambda_2$,
$\mu_2$ define the Liouville wall \cite{DFK}.

Now recall that according to Polchinski's string picture \cite{JPO}
the one
D-instanton amplitude is given by ${A_1}\sim{e^Z}$, where $Z$ is minus
the D-instanton action. From the field theory of instantons in one
dimension \cite{SC} the one instanton amplitude in the semi-classical
approximation to one loop is

\begin{equation}
{A_1}=A{{\left({{\omega}\over{\pi\hbar}}\right)}^{1/2}}
{e^{-{S_0}/\hbar-\omega T/2}}\left[1+O(\hbar)\right],
\end{equation}
where ${S_0}={\int_{-a}^a}dx{{\left(2V\right)}^{1/2}}$ is the instanton
action, $\omega=V''(\pm a)$ and $A$ is a normalisation constant. Here
$V$ is a symmetric double well potencial with two minima at $x=\pm a$,
where $V(\pm a)=0$. In this theory $A_1$ is the one instanton
contribution to the transition amplitude, $<a|{e^{-HT/\hbar}}|-a>$,
for a first quantised particle of unit mass to go from $x=-a$ to $x=a$
as the Euclidean time $t$ goes from $t=-T/2$ to $t=T/2$ for large
$T$. We consider the instanton with its center fixed at $t=0$. In this
picture the instanton is a well localised extended object with a size of
the order $1/\omega$. To connect with the string description we have
to interpret our dual
string coordinate field $Y$ as the Euclidean time $t$. Then the string
has its end points fixed in the center of the instanton. To the disc
level the string can only see a point-like object being sensitive just
to the instanton action. So to tree level both descriptions agree if
$Z=-{S_0}/\hbar$.

In the case of two D-instantons we have seen that at low energy
corresponding to a large distance $\mathcal{Y}$, they interact by
exchanging the massless Klein-Gordon fields. Following Polchinski
\cite{JPO} the two D-instanton amplitude for point-like events is
then

\begin{equation}
{A_2}\sim{e^{2Z+{Z_p}}}.
\end{equation}
Once more from the theory of instantons \cite{SC} the two instanton
amplitude in the semi-classical approximation to one loop is

\begin{equation}
{A_2}={A^2}{{\left({{\omega}\over{\pi\hbar}}\right)}^{1/2}}
{e^{-{S_0}/\hbar-\omega T/2}}\left[1+O(\hbar)\right].
\end{equation}
In the above formula $A_2$ is the two instanton contribution to the
transition amplitude,
$<-a|{e^{-HT/\hbar}}|-a>$, for the particle to go from $x=-a$ to $x=a$
and back to $x=-a$ as time goes from $t=-T/2$ to $t=T/2$, $T\to+\infty$.
In this case the two instantons are in fact an instanton and an
anti-instanton with widely separated centers respectively fixed close
to $t=-T/2$ and to $t=T/2$. In the string picture the instantons are seen
as point-like objects located at $t=-\mathcal{Y}/2$ and at
$t=\mathcal{Y}/2$. They interact with a force ${Z_p}=-\omega
\mathcal{Y}/2$. So to one loop both descriptions agree provided

\begin{equation}
\omega={{\ln(1/{\lambda_2})}\over{\pi\sqrt{\alpha'}}}, \quad
{\lambda_2}>0,\,{\mu_2}\geq 0
\end{equation}
or

\begin{equation}
\omega={{\ln(1/{\mu_2^2})}\over{\pi\sqrt{\alpha'}}}, \quad
{\lambda_2}=0,\,{\mu_2}>0.
\end{equation}
{}From the field theory semi-classical calculation of $A_1$ to one loop
$1/\omega$ is interpreted as the average size of the instanton. Thus
we conclude that the size of the D-instantons is of the order of
$\sqrt{\alpha'}/\ln(1/{\lambda_2})$ for ${\lambda_2}>0$, ${\mu_2}\geq 0$
or of the order of $\sqrt{\alpha'}/\ln(1/{\mu_2^2})$ for ${\lambda_2}=0$,
${\mu_2}>0$.

For the extension to the minimal boundary conformal models we
consider the case $d=1$, $p=0$ where the scalar field
is compactified on a circle of radius $R$. In this case the field is not
single valued when the string is winding around the compact dimension
$Y({\xi^1},{\xi^2})=Y({\xi^1},{\xi^2})+2\pi nR$, $n=1,2,\ldots$. Thus
the periodicity in ${\xi^1}$ can be written as
$Y({\xi^1}+1,{\xi^2})=Y({\xi^1},{\xi^2})+2\pi nR$, which leads us to
$Y({\xi^1},{\xi^2})=2\pi nR{\xi^1}+\bar{Y}({\xi^1},{\xi^2})$,
where $\bar{Y}$ is now a single valued field which in this case
satisfies Neumann boundary conditions. For ${\lambda_2}>0$,
${\mu_2}\geq 0$ the partition function is

\begin{equation}
Z=\ln(1/{\lambda_2}){{\sqrt{\alpha'}}\over{12 R}}.
\end{equation}
This result is to be compared with the partition function on closed
surfaces which agrees with the corresponding matrix model \cite{GK, BK}.
The extra factor of $2$ comes from the open string
Liouville zero mode integral. Also we have naturally lost the self-dual
nature of the closed string. As for the closed surfaces we expect an
agreement with the boundary matrix model. A proof of this however
needs the explicit calculation which is beyond the scope of this work.

For the partition functions with fixed area
$A$ and fixed boundary length $L$ only the integration of the Liouville
zero mode is changed. For $Z(A)$, $Z(L)$ we set ${\lambda_2}=0$,
${\mu_2}=0$ which are the cases where we can
continue to use the DDK scaling argument \cite{MN}. Then

\begin{equation}
Z(A)={1\over{6\alpha A R}},\quad Z(L)={1\over{3\alpha L R}}.
\end{equation}
If as in the case of closed surfaces \cite{BK, DSZ} it could be proved
that the partition functions of the boundary conformal models can be
written as a linear combination of the partition functions of a scalar
field compactified on some particular radii, it would be an easy matter
to write results for these models. Work is in progress in this
direction which we hope to present in a future paper.

If we consider a string which winds around the compact dimension but
still has its end points located on two D-branes a distance $\mathcal{Y}$
apart we
might be able to discuss non-perturbative phenomena in the boundary
conformal models. We then have to take $
Y({\xi^1},{\xi^2})=2\pi n
R{\xi^1}+\mathcal{Y}{\xi^2}-\mathcal{Y}/2+\bar{Y}({\xi^1},{\xi^2})$.
Thus the partition function is

\begin{equation}
Z=-\ln(1/{\lambda_2}) {{\sqrt{\alpha'}}\over{2\pi R}}\ln\left(
1-{e^{-R\mathcal{Y}/\alpha'}}\right).
\end{equation}
As in the Neumann case we also find

\begin{equation}
Z(A)=-{1\over{\pi\alpha
AR}}\ln\left(1-{e^{-R\mathcal{Y}/\alpha'}}\right),\quad
Z(L)=-{2\over{\pi\alpha LR}}\ln\left(1-{e^{-R\mathcal{Y}/\alpha'}}\right).
\end{equation}
For $\mathcal{Y}=0$ we have an extra ultraviolet divergence due to the
winding of the string around the compact dimension.

\section*{Acknowledgments}

We would like to thank Paul Mansfield for valuable discussions.


\begin{thebibliography}{25}
\bibitem[1]{JP} J. Polchinski, Rev. Mod. Phys. 68 (1996) 1245.
\bibitem[2]{CJP} J. Polchinski, ITP preprint NSF-ITP-96-145,
hep-th/9611050 (1996).
\bibitem[3]{DLP} J. Dai, R.G. Leigh and J. Polchinski, Mod. Phys.
Lett. A 4 (1989) 2073; R.G. Leigh, Mod. Phys. Lett. A 4 (1989) 2767.
\bibitem[4]{JPO} J. Polchinski, Phys. Rev. D 50 (1994) R6041.
\bibitem[5]{DDK} F. David, Mod. Phys. Lett. A 3 (1988) 1651; J. Distler
and H. Kawai, Nucl. Phys. B 321 (1989) 509.
\bibitem[6]{MN} P. Mansfield and R. Neves, Nucl. Phys. B 479 (1996) 82.
\bibitem[7]{SS} S. Shenker, in: Random Surfaces and
Quantum Gravity, eds. O. Alvarez, E. Marinari and P. Windey
(Plenum, New York, 1991) p. 191.
\bibitem[8]{ADH} K. Aoki and E. D'Hoker, Mod. Phys. Lett. A 3 (1992)
235; Nucl. Phys. B 490 (1997) 40; Int. J. Mod. Phys. A 12 (1997) 1253.
\bibitem[9]{DFK} P. Di Francesco and D. Kutasov, Nucl. Phys. B 375
(1992) 119.
\bibitem[10]{Bat} A. Erd\'elyi, W. Magnus, F. Oberhettinger and F.G.
Tricomi, Bateman Manuscript Project, Higher
transcendental functions, Vols. 1 and 2 (McGraw-Hill, New York,
1953); Tables of integral
transforms, Vol. 1 (McGraw-Hill, New York, 1954).
\bibitem[11]{LP} J. Liu and J. Polchinski, Phys. Lett. B 203 (1988) 39.
\bibitem[12]{JC} J.L. Cardy, Nucl. Phys. B 240 (1984) 514; Nucl. Phys.
B 324 (1989) 581; J.L. Cardy and D.C. Lewellen, Phys. Lett. B 259 (1991)
274. \bibitem[13]{JS} J. Schulze, Nucl. Phys. B 489 (1997) 580.
\bibitem[14]{CMNP} A. Cohen, G. Moore, P. Nelson and J. Polchinski,
Nucl. Phys. B 281 (1987) 127.
\bibitem[15]{GK} D. Gross and I. Klebanov, Nucl. Phys. B 344 (1990)
475; Nucl. Phys. B 354 (1991) 459.
\bibitem[16]{BK} M. Bershadsky and I. Klebanov, Phys. Rev. Lett. 65
(1990) 3088.
\bibitem[17]{SC} S. Coleman, Aspects of symmetry (Cambridge University
Press, Cambridge, 1985).
\bibitem[18]{DSZ} P. Di Francesco, H. Saleur and J. Zuber, J. Stat. Phys.
49 (1987) 57; V. Pasquier, J. Phys. A 20 (1987) L1229.
\end{thebibliography}
\end{document}